\documentclass[11pt,a4paper]{article}
\pdfoutput=1 

\usepackage[table]{xcolor}
\usepackage{colortbl}
\RequirePackage{ifpdf} 
\usepackage{amsmath} 
\usepackage{mathtools}
\usepackage{cases}
\usepackage{braket}

\usepackage{jheppub}
\usepackage{pstricks}
\usepackage[final]{pdfpages} 
\usepackage{ifpdf} 
\usepackage{slashed}

\usepackage{color}
\usepackage{xcolor}
\definecolor{urlblue}{rgb}{0.2,0.4,0.7}
\definecolor{citegreen}{rgb}{0,0.6,0.2}
\definecolor{linkred}{rgb}{0.9,0.2,0.1}
\usepackage{hyperref}
\hypersetup{
	colorlinks=true, citecolor=citegreen, linkcolor=blue, urlcolor=urlblue}

\usepackage{graphics}
\usepackage{etoolbox} % Load before 'breqn' package to make 'lineno' work
\usepackage{fixmath}
\usepackage{psfrag}

\usepackage{notoccite} % If you have \cite commands in \section-like commands, or in \caption, the
\usepackage{amsfonts}
\usepackage{autobreak}
\usepackage{marginnote}

\title{A Brief Review on the Asymptotic Symmetries of Gravity in Higher Dimensions}

\author{Arpan Kundu$^{~a,b}$}
\affiliation{ $^a$The Institute of Mathematical Sciences, Taramani, Chennai 600113, Tamil Nadu, India.\\
$^b$Homi Bhabha National Institute, Training School Complex, Anushaktinagar, Mumbai	400094, India.}
\abstract{In this brief review we report on the status of asymptotic symmetries of gravity corresponding to the class of metrices named asymptotically flat spacetimes in higher ($d>4$) dimensions. We discuss the consequences of these symmetries both in classical and quantum theories. We also discuss the open issues in these aspects.}

\begin{document}
	\allowdisplaybreaks[4]
	\unitlength1cm
	\maketitle
	\flushbottom
	\newpage
%%%%%%%%%%%%%%%%%%%%%%%%%%%%%%%%%%%%%%%%%%%%%%%%%%%%%%%%%%%%%%%%%%%%%%%%%%%%%%%%%%%%%%%%%%%%%%%%%%%%%%%%%%%%%%%%%%%%%%%%%%%%%%%%%%%%%%%%%%%%%%%%%%%	
\section{Introduction}

Inspired by the seminal work \cite{He:2014laa}, over the last decade there has been a huge progress made in the literature about understanding certain generic properties in the infrared sectors of gauge theories and gravity (see \cite{Strominger:2017zoo} for an early review). These works have established an interesting connection among the following three seemingly disjoint subjects of study: (1) infrared factorisation theorems of the flat-space $\mathcal{S}$-matrix of theory (soft theorems), (2) symmetries that preserve certain large distance behaviour of the fields (asymptotic symmetries) (3) certain physically observable low-frequency effects of radiation (memory effects). Together they have been popularly referred to as Infrared-triangle.

Our focus here will be on gravity. Quest for understanding the properties of quan-
tum gravity has been one of the biggest questions of theoretical high-energy physics. In
the case of spacetimes with a negative cosmological constant (asymptotically Ads space-
times), there exists a fairly well-established notion of holography which goes by the name of
Ads/CFT correspondence (see \cite{Penedones:2016voo} for a recent review). This correspondence holds in any spacetime dimensions. Ads/CFT serves as a theoretical laboratory to explore various
ideas related to quantum gravity in spacetimes with a negative cosmological constant. One
would like to have a similar understanding in the case of spacetime with zero cosmological
constant (asymptotically flat spacetimes). In this aspiration for Flat Holography the
infrared triangle for gravity plays a crucial role.
 
 Research in the last decade has made large progress in understanding the IR-triangle
 of gravity in $d=4$. However, while going to higher dimensions there are additional
 challenges. Earlier works had concluded that the asymptotic symmetries of asymptotically
 flat spacetimes are trivial in higher even dimensions\footnote{In higher odd dimensions, there is a crucial technical roadblock in pursuing similar studies, namely the non-existence of a useful notion of Null Infinity \cite{Hollands:2004ac}. We shall not discuss these issues here and shall restrict to higher even dimensions only.}. But, the Infrared triangle in $d = 4$ and
 the existence of soft theorems in any generic dimensions inspired a bunch of recent works to
 revisit this issue.
 
 In this short review, we summarize this recent progress in understanding the asymptotic symmetries of asymptotically flat spacetimes in higher ($d>4$) even dimensions. The rest of this review is organized as follows. In section-\ref{background}, we introduce the basic background material necessary to understand the rest of the review. In section-\ref{early-ass}, we briefly discuss the early works in asymptotic symmetries in $d=4$ and negative results regarding the existence of non-trivial asymptotic symmetries in higher dimensions. Then in section-\ref{IR-triangle}  we discuss new insights from the Infrared triangle in perturbative quantum gravity in $d=4$ and how they motivate to revisit these negative results. Then, in section-\ref{main-ass} we discuss in detail some of the recent results in understanding the asymptotic symmetries in higher dimensions and how they could bypass the no-go conditions posed by earlier works. We end this short review with a summary and a discussion of the open issues in section-\ref{summary}.
 
 %%%%%%%%%%%%%%%%%%%%%%%%%%%%%%%%%%%%%%%%%%%%%%%%%%%%%%%%%%%%%%%%%%%%%%%%%%%%%%%%%%%%%%%%%%%%%%%%%%%%%%%%%%%%%%%%%%%%%%%%%%%%
 \section{Background}\label{background}

 Asymptotic symmetries are the symmetries of the solution space of a theory preserving certain boundary conditions. In our case, we shall consider gravity theory without cosmological constant coupled with massless fields. This requires preservation of certain beheviour of the metric at it's boundary at infinity. Since, we shall be dealing with the massless fields, we shall be looking at the infinity reached through the null geodesics aka Null Infinity (both in the past $\mathcal{I}^{+}$ and in the future $\mathcal{I}^{-}$). Null infinity ($\mathcal{I}$) in general $d$-dimension has a topology $\mathbb{S}^{(d-2)}\times \mathbb{R}$. We are interested in the class of solutions to Einstein eqn which has some specific behaviour near $\mathcal{I}$, details of which are mentioned later.

 Although, asymptotic symmetries by their very nature is a coordinate independent notion, certain coordinate systems are more useful for various computational purposes. A particularly suitable coordinate system for studying asymptotic symmetries tied to $\mathcal{I}^{+}$ is the retarded Bondi coordinates $(u,r,z)$, where $r$ is the radial distance from the origin, $u=t-r$ is the retarded time and $z$ is the collective coordinate on the conformal sphere $\mathbb{S}^{(d-2)}$. Now, the flat spacetime in this coordinate system can be written as:
 \begin{align}
 	ds^2=-du^{2}-2dudr+r^{2}\gamma_{ab}dz^{a}dz^{b}.
 \end{align}
 Now, $\mathcal{I}^{+}$ in this coordinate system can be reached as $r\rightarrow\infty$ keeping $u$ fixed. The question one asks is that what is the class of all spacetimes that behave like Flat specetime at $\mathcal{I}^{+}$ in a suitable sense? One starts with a general metric in the Bondi gauge :
 \begin{align}\label{bondi-metric}
 	ds^2=  M e^{2\beta} du^2 - 2e^{2\beta} dudr + g_{ab} (dz^a - U^a du) (dz^b - U^b du),
 \end{align}
 where the Bondi gauge condition is given by
 \begin{align}\label{bondi-gauge}
 	g_{rr}=0\quad g_{ra}=0 \quad \det\Big(\frac{g_{ab}}{r^2} \Big)= \det(\gamma_{ab}).
 \end{align}
 Here, $M(u,r,z)$, $U^{a}(u,r,z)$, $\beta(u,r,z)$ and $g_{ab}(u,r,z)$ are parameters, which are general functions of the coordinates. Asymptotic flatness is ensured by demanding preservation of certain large-$r$ fall-off of the metric near $\mathcal{I}^{+}$. This requires putting certain large $r$ behaviour of these functions. There is no unique condition for this, but one is guided by the following basic principles: (1) the conditions should be weak enough to allow physically interesting solutions like black holes, and  gravitational waves; (2) the conditions should be strong enough to ensure that physical quantities like charge, Mass, Angular Momentum, etc. don't diverge. The algebra of non-trivial symmetry transformations which preserve a specific boundary condition, is called the asymptotic symmetry algebra (ASA). Since there is no unique fall-off condition for AFS, there is no unique ASA. Weakening the fall-off corresponds to the enlargement of the ASA. 
%%%%%%%%%%%%%%%%%%%%%%%%%%%%%%%%%%%%%%%%%%%%%%%%%%%%%%%%%%%%%%%%%%%%%%%%%%%%%%%%%%%%%%%%%%%%%%%%%%%%%%%%%%%%%%%%%%%%%%
%%%%%%%%%%%%%%%%%%%%%%%%%%%%%%%%%%%%%%%%%%%%%%%%%%%%%%%%%%%%%%%%%%%%%%%%%%%%%%%%%%%%%%%%%%%%%%%%%

\section{Early Works on Asymptotic Symmetries in Gravity}\label{early-ass}

 Study of asymptotic symmetries can be traced back to as early as sixties. In the seminal works \cite{Bondi:1962px,Sachs:1962wk}, in $d=4$, the ASA was obtained to be the celebrated $\mathfrak{BMS}$ algebra, which is a semidirect product of Supertranslation ($\mathfrak{ST}$) and Lorentz. $\mathfrak{ST}$ itself is an infinite dimensional enlargement of the translation subalgebra of the Poincare algebra (which is the semi-direct product of Translation and Lorentz). $\mathfrak{ST}$ algebra is parametrized by a free function $f(z)$ on $\mathbb{S}^{2}$. In $d=4$, this $\mathfrak{BMS}$ algebra was further extended in later works. In $d=4$, there are different proposals for infinite dimensional extension of the $\mathfrak{BMS}$ algebra, using different infinite-dimensional extensions of the Lorentz subalgebra of the $\mathfrak{BMS}~(\mathfrak{ST}\ltimes\mathrm{Lorentz})$ algebra. In $d=4$, Lorentz transformation induces the global conformal transformations on $\mathbb{S}^{2}$. Inspired by an attempt to build a proposed BMS-CFT correspondence (in analogy to Ads/CFT), in  the Extended BMS ($\mathfrak{EBMS}$) proposal \cite{Barnich:2009se,Barnich:2010eb}, the $\mathrm{Lorentz}$ algebra is extended to include local conformal transformations on $\mathbb{S}^{2}$. Later, inspired by an attempt to build an improved understanding of Infrared Triangle \footnote{We shall return to this point in a bit more detail in the next section-\ref{IR-triangle}.}, in the Generalised BMS ($\mathfrak{GBMS}$) proposal \cite{Campiglia:2015yka}, the $\mathrm{Lorentz}$ algebra is extended to include any area preserving smooth diffeomorphisms of $\mathbb{S}^{2}$. Both of these infinite dimensional extensions of the Lorentz algebra are called Superrotation in $d=4$. In the $\mathfrak{EBMS}$ case, Superrotation is parametrized by holomorphic vector fields $V^{a}(z)$ on $\mathbb{S}^{2}$, whereas in the $\mathfrak{GBMS}$ case, Superrotation is parametrized by any smooth area preserving vector fields $V^{a}(z)$ on $\mathbb{S}^{2}$. It is important to keep in mind that, $\mathfrak{ST}$ algebra is a subalgebra of all three proposed ASA in $d=4$, namely $\mathfrak{BMS}$, $\mathfrak{EBMS}$, and $\mathfrak{GBMS}$.

   In \cite{Hollands:2016oma}, the ASA corresponding to AFS of even $d\geq4$ was studied in their connection to displacement memory effect. The displacement memory effect is the DC shift observed in the pair of gravity wave detectors due to the passing of a burst gravitational wave. In \cite{Hollands:2016oma}, it was argued that, in $d=4$ the Supertranslations are tied to the displacement memory effect, and if one uses a strict fall-off such that the ASA is Poincare and thus disallowing Supertranslations, generic radiative solutions are automatically excluded. Hence, allowing Supertranslation is essential in $d=4$. In contrast, in $d>4$, while the memory effects are seen corresponding to $\mathcal{O}(r)$ at the $r$-expansion of the angular part of the metric, radiation occurs at $\mathcal{O}(r^{-(d-2)})$.  Hence, enlargement of the Poincare algebra to include Supertranslation doesn't become a physical necessity. Furthermore, allowing for Supertranslation leads to divergent physical quantities. In this logic, it was argued that the Supertranslation doesn't exist in even dimension $d>4$. (See \cite{Hollands:2003ie,Tanabe:2011es} for earlier works regarding asymptotic symmetries of higher dimensional gravity, which also gives negative results about the existence of Supertranslation in higher even dimensions.) However, certain new insights from the study of $\mathcal{S}$-matrix in the corresponding quantum theory in $d=4$ have led to revisiting the ASA in higher d. We shall discuss these motivations in section-\ref{IR-triangle}.
%%%%%%%%%%%%%%%%%%%%%%%%%%%%%%%%%%%%%%%%%%%%%%%%%%%%%%%%%%%%%%%%%%%%%%%%%%%%%%%%%%%%%%%%%%%%%%%%%%%%%%%%%%%%%%%%%%%%%%%%%%%%%%%%%%%%%%%
\section{New Insights from Infrared Triangle}\label{IR-triangle}
So far we have talked about asymptotic symmetries only in classical theory. One can ask what are the implications of these symmetries at the level of quantum gravity. More specifically, can we say anything about the properties of perturbative quantum gravity $\mathcal{S}$-matrix? Starting with \cite{He:2014laa}, a program was initiated in which certain already known factorisation theorems of perturbative quantum gravity $\mathcal{S}$-matrix have been found to be a consequence of elevating the asymptotic symmetries of the classical theory as a conjectured symmetry of the $\mathcal{S}$-matrix of the corresponding quantum theory. These factorisation theorems are called Soft graviton theorems\footnote{Soft Theorems hold for any gauge theory, but our focus here will be on soft gravitons only.}.

 Consider a scattering amplitude containing finite energy particles of any mass, spin, and one  soft (energy $\omega$$\rightarrow$0) graviton. The amplitude can be factorised in terms of the amplitude of other finite energy particles without the soft graviton and some universal factor called the Soft factor. In fact, there are similar factorisation theorems for more than one soft gravitons, but we shall focus in this article mainly on Single Soft Graviton theorems.
    
Early works on the Soft graviton theorem in tree-level perturbative quantum gravity can be traced back to the sixties \cite{Weinberg:1965nx}. Later, in \cite{Cachazo:2014fwa}, Soft graviton theorems were extended to subleading orders in the energy of the Soft graviton. Recent works by Sen and his collaborators \cite{Sen:2017xjn,Sen:2017nim,Laddha:2017ygw,Chakrabarti:2017ltl} have put this on a much more robust footing by proving soft graviton theorems for arbitrary but finite number of soft gravitons in any generic theories of quantum gravity in generic dimensions where finite energy particles can have any mass and spin. In $d\geq5$ due to the absence of infrared divergence, these factorisation theorems are true for all-loop amplitudes. In $d=4$ infrared divergences force to make these statements at the tree-level and there are additional logarithmic corrections \cite{Sahoo:2018lxl} to the soft factors once the loop effects are taken into account.

 In this section, we first introduce Leading and Subleading Soft graviton theorems in general dimensions and discuss how in $d=4$ they are related to the asymptotic symmmetries. Then, we discuss the early hints that showed similar relations might hold in higher even dimensions as well.  
 
%%%%%%%%%%%%%%%%%%%%%%%%%%%%%%%%%%%%%%%%%%%%%%%%%%%%%%%%%%%%%%%%%%%%%%%%%%%%%%%%%%%%%%%%%%%%
\subsection{Leading Soft Theorem \& Supertranslation Symmetry in $d=4$}

We want to briefly revisit how Leading Soft Graviton Theorem is related to the conjectured Supertranslation Symmetry of the quantum garvity $\mathcal{S}$-matrix. Let us start by stating the Leading soft graviton theorem \cite{Weinberg:1965nx,Sen:2017xjn} first.
 
Consider a scattering amplitude containing $i=1,\cdots,n$ finite energy particles of any mass, spin and one soft graviton (energy going to zero in the limiting sense). Then the Leading Soft Graviton Theorem can be written as:

\begin{align}\label{leading-ST}
	\lim\limits_{\omega \rightarrow 0} \omega\bra{\mathrm{Out}}\mathfrak{a}_{\lambda}(\omega,z_s)\mathcal{S}\ket{\mathrm{in}}=\sqrt{8\pi G_{N}}\Bigg(\sum_{i}\frac{\epsilon_{\lambda}^{\mu\nu}k^i_{\mu}k^{i}_{\nu} }{(p/\omega)\cdot k^i }\Bigg)	\bra{\mathrm{out}}\mathcal{S}\ket{\mathrm{in}}.
\end{align}	

Here, $p^{\mu}$ and $\epsilon_{\mu\nu}$ are the momentum and polarisation tensor of the soft graviton with polarisation label $\lambda$. $\mathfrak{a}_{\lambda}(\omega,z_s)$ creates a soft graviton in the ``Out" state with energy $\omega$ whose direction on the celestial sphere can be denoted using collective coordinate $z_{s}$. $k^{i}_{\mu}$ is the momentum of the $i$-th finite energy particle.

Although, the connection between soft theorem and asymptotic symmetry can be build for finite energy particles with any mass and spins, let us now restrict to perturbative gravity coupled to a massless scalar for simplicity. In this case, it was shown in \cite{He:2014laa} that the above soft theorem \eqref{leading-ST} is a consequence of the conjectured Supertranslation symmetry of the $\mathcal{S}$ matrix. 

One can derive this equivalence starting from the Soft theorem and then derive a Ward
identity of Supertranslation for the $\mathcal{S}$-matrix. In this way, one obtains a Ward identity of
the form
\begin{align}\label{st-ward-4d}
	\bra{\mathrm{Out}}[\mathcal{Q}^{d=4}_{\mathrm{ST}},\mathcal{S}]\ket{\mathrm{in}}=0,
\end{align}
where, $\mathcal{Q}^{d=4}_{\mathrm{ST}}$ is the quantized version of the Supertranslation charge in $d=4$. Now, since the charge obtained from the Soft theorem matches with the charge obtained from classical gravity this proves that the Soft theorem \eqref{leading-ST} is a consequence of the Supertranslation Symmetry.

Another way is to start from the classical symmetry and obtain a conserved charge ($\mathcal{Q}^{d=4}_{\mathrm{ST}}$). The charges are parametrized by free function $f(z)$ on $\mathbb{S}^{2}$. Then elevate this classical symmetry to the symmetry of the quantum gravity $\mathcal{S}$-matrix by writing a Ward identity of Supertransaltion \eqref{st-ward-4d}. Finally, from this one derives the soft theorem \eqref{leading-ST} as a consequence of the Ward identity \eqref{st-ward-4d}. 

A few conceptual points need to be stated here.  Apriori there are two independent $\mathfrak{BMS}$ algebras: (1) $\mathfrak{BMS}^{+}$ acting on $\mathcal{I}^{+}$, labelled by free function $f^{+}(z)$ and (2) $\mathfrak{BMS}^{-}$ acting on $\mathcal{I}^{-}$, labelled by free function $f^{-}(z)$. In \cite{Strominger:2013jfa}, a diagonal subalgebra $\mathfrak{BMS}^{0}$ of $\mathfrak{BMS}^{+}\times\mathfrak{BMS}^{-}$ was identified as the symmetry of the gravitational scattering problem. This is done through the  antipodal matching $f^{+}(z)=f^{-}(-z)$. Also, while going from the Ward identity \eqref{st-ward-4d} to soft theorem \eqref{leading-ST} one needs to choose the free function $f(z)$ such that it localises on the particular direction of the soft graviton. Hence, Leading Soft Theorem can be thought of as a consequence of Spontaneous Supertranslation Symmetry Breaking in the space of degenerate vacua. 

It is also worth mentioning here that the Supertranslation symmetry in $d=4$ is related to classical observable effects called gravitational displacement memory \cite{Strominger:2014pwa}.

%%%%%%%%%%%%%%%%%%%%%%%%%%%%%%%%%%%%%%%%%%%%%%%%%%%%%%%%%%%%%%%%%%%%%%%%%%%%%%%%%%%%%%%%%%%%%%%%%%%%%%%%%%%%%%%%%%
\subsection{Subleading Soft Graviton Theorem \& Superrotation Symmetry in $d=4$}
Soft factorization of the amplitude holds at the subleading level of the energy of the soft graviton as well \cite{Cachazo:2014fwa}. Consider a scattering amplitude containing $i=1,\cdots,n$ finite energy particles of any mass, spin and one soft graviton (energy going to zero in the limiting sense).  Then the Subleading Soft Graviton Theorem can be written as:

\begin{align}\label{subleading-ST}
	\lim\limits_{\omega \rightarrow 0}(1 + \omega\partial_{\omega})\bra{\mathrm{Out}}\mathfrak{a}_{\lambda}(\omega,z_s)\mathcal{S}\ket{\mathrm{in}} =-i\sqrt{8\pi G_{N}}\Bigg(\sum_{i}\frac{\epsilon_{\lambda}^{\mu\nu}k^i_{\nu}p^{\rho}\mathcal{J}^i_{\mu\rho} }{p\cdot k^i }\Bigg)\bra{\mathrm{out}}\mathcal{S}\ket{\mathrm{in}}.
\end{align}

Here, as before $p^{\mu}$ and $\epsilon_{\mu\nu}$ are the momentum and polarisation tensor of the soft graviton with polarisation label $\lambda$. $\mathfrak{a}_{\lambda}(\omega,z_s)$ creates a soft graviton in the ``Out" state with energy $\omega$ whose direction on the celestial sphere can be denoted using collective coordinate $z_{s}$. $\mathcal{J}^i_{\mu\nu}$ is the angular momentum of the $i$-th finite energy particle.

As before, for simplicity let us restrict to gravity coupled to a massless scalar. One wants to ask like the leading case, in the subleading case whether there is an asymptotic symmetry origin of such factorisation. In \cite{Kapec:2014opa}, in $d=4$, a Starting from the Subleaing Soft Theorem  a Ward identity of the form  
\begin{align}\label{sr-ward-4d}
	\bra{\mathrm{Out}}[\mathcal{Q}^{d=4}_{\mathrm{SR}},\mathcal{S}]\ket{\mathrm{in}}=0
\end{align}
was derived, where, $\mathcal{Q}^{d=4}_{\mathrm{SR}}$ is the quantized version of the Superrotation charge in $d=4$ corresponding to $\mathfrak{EBMS}$ algebra. However, the singular nature of the vector fields restricted the proving of this equivalence in the other way around, namely, Ward identity \eqref{sr-ward-4d} to Soft theorem \eqref{subleading-ST}. This prompted the authors of \cite{Campiglia:2014yka} to propose a different definition of Superrotation based on Diff$(\mathbb{S}^{2})$ vector field as mentioned in section-\ref{early-ass}. This corresponds to the proposal of $\mathfrak{GBMS}$ algebra as the ASA for AFS in $d=4$. In the case of Superrotations corresponding to $\mathfrak{GBMS}$ one can go both-ways: from Ward identity \eqref{sr-ward-4d} to Soft theorem \eqref{subleading-ST} and the reverse. A first principle derivation of the charges was given in \cite{Campiglia:2015yka}. 

Like the leading case, a few conceptual points need to be stated here.  Apriori there are two independent $\mathfrak{GBMS}$ algebras: (1) $\mathfrak{GBMS}^{+}$ acting on $\mathcal{I}^{+}$, labelled by free functions ($f^{+}(z)$, $V^{a}_{+}(z)$) and (2) $\mathfrak{GBMS}^{-}$ acting on $\mathcal{I}^{-}$, labelled by free function ($f^{-}(z)$, $V^{a}_{-}(z)$). Insoired from \cite{Strominger:2013jfa}, a diagonal subalgebra $\mathfrak{GBMS}^{0}$ of $\mathfrak{GBMS}^{+}\times\mathfrak{GBMS}^{-}$ can identified as the symmetry of the gravitational scattering problem. This is done through the following antipodal matching
 \begin{align}
 	f^{+}(z)=f^{-}(-z)\quad\qquad V^{a}_{+}(z)=V^{a}_{-}(-z).
 \end{align}
  Also, while going from the Ward identity \eqref{sr-ward-4d} to soft theorem \eqref{subleading-ST} one needs to choose the free vector fields $V^{a}(z)$ such that it localises on the particular direction of the soft graviton. Hence, Subeading Soft Theorem can be thought of as a consequence of Spontaneous Superrotation Symmetry breaking in the space of degenerate vacua. This corresponds to spontaneous symmetry breaking from $\mathfrak{GBMS}$ to $\mathfrak{BMS}$.
  
   It is also worth mentioning here that the Superrotation symmetry in $d=4$ is related to classical observable effects called the gravitational Spin memory \cite{Pasterski:2015tva}.
%%%%%%%%%%%%%%%%%%%%%%%%%%%%%%%%%%%%%%%%%%%%%%%%%%%%%%%%%%%%%%%%%%%%%%%%%%%%%%%%%%%%%%%%%%%%%%%%%%%%%%%%%%%%%%%%
\subsection{Ward identities from Soft Theorem in Higher Even dimesions}
In $d=4$ Leading Soft Graviton Theorem follows from the supertranslation symmetry of the $\mathcal{S}$-matrix \cite{He:2014laa}. Since the leading soft graviton theorem \eqref{leading-ST} holds in all dimensions, a natural question is whether supertranslation also exists in all dimensions. Contrary to the classical result of \cite{Hollands:2016oma}, in \cite{Kapec:2015vwa}, based on the factorisation properties of the perturbative quantum gravity $\mathcal{S}$-matrix, it was argued that the Supertranslation (and correspondingly $\mathfrak{BMS}$) holds even in higher even ($d=2m+2$) dimension and a Supertranslation compatible fall-offs of the Bondi metric \eqref{bondi-metric} were proposed.  In \cite{Kapec:2015vwa}, in all higher even dimensions a Ward identity for the $\mathcal{S}$-matrix of the following form was derived starting from the Leading Soft Graviton Theorem \eqref{leading-ST}.
\begin{align}
		\bra{\mathrm{Out}}[\mathcal{Q}^{d=2m+2}_{\mathrm{ST}},\mathcal{S}]\ket{\mathrm{in}}=0
\end{align}

 From this Ward identity, Supertranslation charge ($\mathcal{Q}^{d=2m+2}_{\mathrm{ST}}$) can be read-off in generic higher even dimension. This charge was shown to generate the Supertranslation using some proposed commutation relation among the radiative degrees of freedom. However, since there was no first principle derivation of the charge in classical gravity, this created an apparent contradiction with the results of classical gravity \cite{Hollands:2016oma} , the resolution of which will be discussed in the next section.
 
 Inspired from \cite{Kapec:2015vwa}, in \cite{Colferai:2020rte}, based on an attempted generalisation of Diff($\mathbb{S}^{2}$) Superrotation in higher even dimensions (in terms of Diff($\mathbb{S}^{2m}$) vector fields) a Ward identity of Superrotation of the following form was derived in linearized gravity in higher even dimensions starting from Subleading Soft Graviton Theorem \eqref{subleading-ST}:
\begin{align}
	\bra{\mathrm{Out}}[\mathcal{Q}^{d=2m+2}_{\mathrm{SR}},\mathcal{S}]\ket{\mathrm{in}}=0.
\end{align}
However, lacking a first principle understanding of the $\mathfrak{GBMS}$ symmetries in higher dimensions in classical gravity, it wasn't clear whether one can indeed generalise superrotations in higher dimensions in terms of Diff($\mathbb{S}^{2m}$) vector fields and whether one can properly embed $\mathfrak{BMS}$ algebra as a subalgebra of this $\mathfrak{GBMS}$ algebra. This issue was addressed in later works \cite{Chowdhury:2022gib,Capone:2023roc}.
%%%%%%%%%%%%%%%%%%%%%%%%%%%%%%%%%%%%%%%%%%%%%%%%%%%%%%%%%%%%%%%%%%%%%%%%%%%%%%%%%%%%%%%%%%%%%%%%%%%%%%%%%%%%%%%%%%%%%%%%%%%%%%%%%%%%%%%%%%%%%%%%%

\section{Revisiting the Asymptotic Symmetries in Higher Even Dimensions}\label{main-ass}
As already mentioned, regarding the non-trivial ASA in higher even dimensions, there is a contradiction between the results obtained from classical gravity \cite{Hollands:2016oma} and from the factorisation property of  quantum gravity $\mathcal{S}$-matrix \cite{Kapec:2015vwa}. This apparent contradiction was resolved in \cite{Aggarwal:2018ilg}. The author made a derivation of Supertranslation charge in linearized gravity in higher even dimensions using the Covariant Phase Space Formalism \cite{Compere:2018aar}. Despite having the fall-off conditions that allow for Supertranslations, the author was able to get a finite charge by adding certain additional boundary conditions at the boundaries of the $\mathcal{I}^{+}$ and hence, bypassing the no-go conditions of \cite{Hollands:2016oma}. Interestingly, these additional conditions also ensure the correct counting for the number of independent soft theorems. Hence it established the existence of Supertranslation in the higher even dimensions on a stronger footing.

The analysis of \cite{Aggarwal:2018ilg} was further strengthened in favour of the existence of Supertranslation in higher even dimensions in \cite{Chowdhury:2022nus}, where the authors did the covariant phase space analysis in non-linear gravity focussing on $d=6$.

In the following, we first summarize lessons from the above results in a more concrete manner. Then we discuss how one can generalise $\mathfrak{BMS}$ to $\mathfrak{GBMS}$ in $d=6$ and discuss its consequences.
%%%%%%%%%%%%%%%%%%%%%%%%%%%%%%%%%%%%%%%%%%%%%%%%%%%%%%%%%%%%%%%%%%%%%%%%%%%%%%%%%%%%%%%%%%%%%%%%%%%
\subsection{Supertranslations in Higher Even Dimensions \& Consequences}\label{ST}

In \cite{Kapec:2015vwa,Aggarwal:2018ilg} the analysis was done in the linearized gravity coupled to matter, and hence, the authors worked with the linearized Bondi metric, which can be written as:
\begin{align}
	ds^2=  M  du^2 - 2 dudr + g_{ab}dz^{a}dz^{b}-2 U_{a}dz^a du.
\end{align}

The fall-off conditions choosen for the parameters were:
\begin{align}\label{linear-BMS-fall-off-hd}
	&M = -1+\sum\limits_{n = 1}^{\infty} \frac{M^{(n)}(u, z)}{r^n},\quad  
	U_a = \sum\limits_{n = 0}^\infty \frac{U_a^{(n)}(u, z)}{r^n}\nonumber\\
	&g_{ab} = r^2 \gamma_{ab}(z) + \sum\limits_{n = -1}^\infty \frac{C_{ab}^{(n)}(u, z)}{r^n}
\end{align}
In linear theory, the determinant condition in \eqref{bondi-gauge} ensures that $\gamma^{ab}C_{ab}^{(n)}=0 ~\forall~ n$, i.e. all $C_{ab}^{(n)}$ are traceless. From the Einstein's eqn one can show that $\partial_{u}C^{(-1)}_{ab}=0$ and $C^{(m-2)}_{ab}$ is the free radiative data. Supertranslations are generated by the vector fields:
\begin{align}\label{ST-vec-field}
	\xi_{\mathrm{ST}}=f(z)\partial_{u}-\gamma^{ab}(z)\mathcal{D}_{a}f(z)\partial_{b}+\frac{1}{2m}\mathcal{D}^{2}f(z)\partial_{r}+\cdots
\end{align}  
Here, $f(z)$ is any smooth function on $\mathbb{S}^{(d-2)}$, and $\cdots$ denotes the subleading (in $r$) orders  of the vector fields. The action of Supertranslation preserves the fall-off \eqref{linear-BMS-fall-off-hd} and thus Supertranslation qualifies as a valid candidate for asymptotic symmetry provided one gets a finite non-zero Noether charge corresponding to it. 

 Supertranslation does a shift of the $C^{(-1)}_{ab}$ as :
\begin{align}
	\delta_{\mathrm{ST}}C^{(-1)}_{ab}=\frac{1}{m}\mathcal{D}^{2}f\gamma_{ab}-2\mathcal{D}_{a}\mathcal{D}_{b}f
\end{align}
In the linearized theory,  $\delta_{\mathrm{ST}}C_{ab}^{(n)}=0 ~\forall~ n\geq0$ (including the radiative order $m-2$). However, later we shall see that this isn't true for non-linear gravity and supertranslation indeed does affect the radiative order as well. 

Using the covariant-phase space techniques (for review see \cite{Compere:2018aar}), in \cite{Aggarwal:2018ilg}, the Noether charge was calculated for general even dimension $d=2m+2$. Since the analysis was done in the linearized gravity the hard part \footnote{similar to the $d=4$ case, the nomenclature soft and hard is used to denote the part of the charge linear in the gravitational free data and quadratic in the gravity/matter free data respectively.} of the charge $\mathcal{Q}^{\mathrm{Hard,Lin}}_{\mathrm{ST}}=\int_{\mathcal{I}^{+}}f(z)\mathcal{T}^{\mathrm{Matter}(2m)}_{uu}$ doesn't contain any contribution from the gravitational free data and depends on the matter only. Here,$\mathcal{T}^{\mathrm{Matter}(2m)}_{uu}$ stands for the term at the $r^{-2m}$ order in the large-$r$ expansion of $uu$ component of the matter stress energy tensor. 

The soft charge contained finite as well as the divergent term. Divergence could be cured by putting certain additional $2m-2$ conditions on the behaviour of $C^{(n)}_{ab}$ s at the boundaries of the $\mathcal{I}^{+}$. These conditions are:
\begin{align}\label{finite-charge-condition-hd}
	&\mathcal{D}^{a}\mathcal{D}^{b}C_{ab}^{(n)}=u^{n+1}\Big[\prod^{n}_{j=0}\mathfrak{D}_{j,m}\Big]\mathcal{D}^{a}\mathcal{D}^{b}C_{ab}^{(-1)}\quad\forall~~ 0\leq n\leq m-3\nonumber\\
	&\mathcal{D}^{a}\mathcal{D}^{b}C_{ab}^{(m-2)}\Bigg|_{u=\pm\infty,z}\sim\mathcal{O}(|u|^{-m+1-\epsilon})\quad\epsilon>0\nonumber\\
	&\mathcal{D}^{a}\mathcal{D}^{b}C_{ab}^{(m+n-2)}\Bigg|_{u=\pm\infty,z}\sim\mathcal{O}(|u|^{-m+1+n-\epsilon})\quad \forall~~1\leq n\leq m-2,~\epsilon>0,\nonumber\\
\end{align} 
where,
\begin{align}
	\mathfrak{D}_{j,m}=\frac{j(2m-j-3)}{2(j+2)(-2m+j+1)(-m+j+2)}\Big[\mathcal{D}^{2}-(j+1)(2m-j-2)\Big].
\end{align}
In \cite{Aggarwal:2018ilg}, a aposteriori motivation for putting these conditions were given. It is important to note that, in the $d$ dimension, there are $d(d-3)/2$ leading soft theorems corresponding to the number of polarisations of the graviton. However, all of them are not independent. Since, we have only one soft charge, and in that only one free function to choose, this means, there is only one independent soft theorem. This means one needs $d(d-3)/2-1$ extra conditions. These conditions are called the ``Generalised CK conditions" in higher dimensions, in analogy with the Christodoulou-Klainerman (CK) condition in $d=4$ \cite{Christodoulou:1993uv,Strominger:2013jfa}, which give the correct counting for the number of independent soft theorems. Among these $d(d-3)/2-1$ conditions $(d-4)=(2m-2)$ conditions, are the conditions \eqref{finite-charge-condition-hd} necessary for the finiteness of charge \cite{Aggarwal:2018ilg}. The remaining $(d-2)(d-3)/2$ other conditions $D_{a}U^{(0)}_{b}=D_{b}U^{(0)}_{a}$ are obtained from the vanishing of magnetic part of the Weyl tensor at $\mathcal{O}(r^{-1})$ \cite{Kapec:2015vwa}.

The finite part of the soft charge obtained in \cite{Aggarwal:2018ilg} mathched with \cite{Kapec:2015vwa}, where it was derived from the soft theorems. This finite Soft charge is given by:
\begin{align}
&\mathcal{Q}^{\mathrm{Soft,Lin}}_{\mathrm{ST}}\nonumber\\

&=\frac{1}{8\pi G_{N}} \frac{1}{(2m-1)}\frac{2^{-m}}{\Gamma(m)}\int_{\mathcal{I}^{+}}f(z)\prod_{l=m+1}^{2m-1}\Big[\mathcal{D}^{2}-(2m-l)(l-1)\Big]I^{(m-2)}\Big(\mathcal{D}^{a}\mathcal{D}^{b}C_{ab}^{(m-2)}\Big)	
\end{align}
where the operator $I^{(n)}$ stands for is $n$-th antiderivative of the argument with respect to $u$ i.e. $I^{(n)}=[\int_{u}]^{n}$.  Note that, $\int_{\mathcal{I}^{+}}=\int d^{2m}z\sqrt{\gamma}\int_{u}$ and the $\int_{u}I^{(m-2)}\Big(\mathcal{D}^{a}\mathcal{D}^{b}C_{ab}^{(m-2)}\Big)	$ gives the zero mode.

Total supertranslation charge in linearized gravity in any general higher even dimension $d=2m+2$ is thus given by:
\begin{align}\label{ST-charge-general-d}
\mathcal{Q}^{\mathrm{Lin}}_{\mathrm{ST}}=&\mathcal{Q}^{\mathrm{Soft,Lin}}_{\mathrm{ST}}+\mathcal{Q}^{\mathrm{Hard,Lin}}_{\mathrm{ST}}\nonumber\\
&=\frac{1}{8\pi G_{N}}\frac{1}{(2m-1)}\frac{2^{-m}}{\Gamma(m)}\int_{\mathcal{I}^{+}}f(z)\prod_{l=m+1}^{2m-1}\Big[\mathcal{D}^{2}-(2m-l)(l-1)\Big]I^{(m-2)}\Big(\mathcal{D}^{a}\mathcal{D}^{b}C_{ab}^{(m-2)}\Big)\nonumber\\
&\quad\quad\quad\quad\quad\quad\quad\quad\quad\quad+\frac{1}{8\pi G_{N}}	\int_{\mathcal{I}^{+}}f(z)\mathcal{T}^{\mathrm{Matter}(2m)}_{uu}
\end{align}
In $d=4$, supertranslation charge obtained from the covariant phase space analysis matches with the ``electric charge" obtained from the Weyl tensor \cite{Campiglia:2015yka}. In \cite{Aggarwal:2018ilg}, it was shown that the same is true for higher even dimension as well, since the charge \eqref{ST-charge-general-d} is the same as the ``electric charge" ($\mathcal{Q}^{\mathrm{Elec}}[\xi_{\mathrm{ST}}]$) obtained from the Weyl tensor:
\begin{align}
\mathcal{Q}^{\mathrm{Lin}}_{\mathrm{ST}}=\mathcal{Q}^{\mathrm{Elec}}[\xi_{\mathrm{ST}}]\equiv-\frac{1}{8\pi G_{N}}\frac{1}{2m-1}\lim_{t\rightarrow\infty}\int_{\Sigma_{t}}\partial_{\mu}\Big[r\sqrt{g}C^{\mu t}_{~~\lambda r}\xi^{\lambda}_{\mathrm{ST}}\Big]
\end{align}
where, $\xi_{\mathrm{ST}}$ is the supertranslation vector field and $C_{\mu\nu\rho\sigma}$ is the Weyl tensor.

So far, we have talked about the generic higher even dimensions. Let us now focus on the results of \cite{Aggarwal:2018ilg} in $d=6$ in particular, as we will discuss this case in detail for non-linear gravity. For notational ease, in $d=6$, we shall denote $C^{(0)}_{ab}$ as $D_{ab}$ and $C^{(-1)}_{ab}$ as $C_{ab}$. Here, $D_{ab}(u,z)$ is the dynamical mode and $C_{ab}(z)$ is the pure supertranslation mode. Higher $C^{(n)}_{ab}$'s will not be important for discussion in $d=6$ as they don't contribute at $\mathcal{I}^{+}$. The supertranslation soft charge in $d=6$ has a finite and a divergent piece. The divergence is cured by imposing the following $u$ fall-off of the dynamical mode at the boundary of the $\mathcal{I}^{+}$:
\begin{align}\label{finite-condition-6d}
	\mathcal{D}^{a}\mathcal{D}^{b}D_{ab}(u=-\infty,z)=\mathcal{D}^{a}\mathcal{D}^{b}D_{ab}(u=+\infty,z)=\mathcal{O}(|u|^{-1-\epsilon}) ,\quad\epsilon>0.
\end{align}
Finally, soft supertranslation charge is given by: 
\begin{align}\label{Soft-ST-6d}
	\mathcal{Q}^{\mathrm{Soft,Lin}}_{\mathrm{ST}}=\frac{1}{96\pi G_{N}} \int_{\mathcal{I}^{+}}f(z)(\mathcal{D}^{2}-2)\mathcal{D}^{a}\mathcal{D}^{b}D_{ab}=\frac{1}{96\pi G_{N}}\int_{\mathbb{S}^{4}}f(z)(\mathcal{D}^{2}-2)\mathcal{D}^{a}\mathcal{D}^{b}\mathcal{N}^{(0)}_{ab},
\end{align}
where $\mathcal{N}^{(0)}_{ab}=\int_{u}D_{ab}$ is the leading soft mode.

Hard supertranslation charge is given by:
\begin{align}\label{hard-ST-6d}
	\mathcal{Q}^{\mathrm{Hard,Lin}}_{\mathrm{ST}}=\frac{1}{8\pi G_{N}}\int_{\mathcal{I}^{+}}f(z)\mathcal{T}^{\mathrm{Matter}(4)}_{uu}
\end{align}

Finally, one can write the total supertranslation charge in linearized gravity in $d=6$ as \cite{Aggarwal:2018ilg}:
\begin{align}\label{ST-charge-lin}
\mathcal{Q}^{\mathrm{Lin}}_{\mathrm{ST}}=&\mathcal{Q}^{\mathrm{Soft,Lin}}_{\mathrm{ST}}+\mathcal{Q}^{\mathrm{Hard,Lin}}_{\mathrm{ST}}\nonumber\\
&=\frac{1}{96\pi G_{N}}\int_{\mathcal{I}^{+}}f(z)(\mathcal{D}^{2}-2)\mathcal{D}^{a}\mathcal{D}^{b}D_{ab}+\frac{1}{8\pi G_{N}}\int_{\mathcal{I}^{+}}f(z)\mathcal{T}^{\mathrm{Matter}(4)}_{uu}\nonumber\\	
&=\frac{1}{96\pi G_{N}}\int_{\mathbb{S}^{4}}f(z)(\mathcal{D}^{2}-2)\mathcal{D}^{a}\mathcal{D}^{b}\mathcal{N}^{(0)}_{ab}+\frac{1}{8\pi G_{N}}\int_{\mathcal{I}^{+}}f(z)\mathcal{T}^{\mathrm{Matter}(4)}_{uu}	
\end{align}

So far, we have talked about asymptotically flat spacetime in linearised gravity.  In \cite{Chowdhury:2022nus}, the work of \cite{Aggarwal:2018ilg} was extended to non-linear gravity focussing on $d=6$. One starts with the general metric \eqref{bondi-metric} satisfying the Bondi gauge \eqref{bondi-gauge} and impose the following fall-off condition:
\begin{align}\label{BMS-fall-off-hd}
	&M = -1+\sum\limits_{n = 1}^{\infty} \frac{M^{(n)}(u, z)}{r^n},\quad  
	\beta = \sum\limits_{n = 2}^\infty \frac{\beta^{(n)}(u, z)}{r^n},\quad
	U_a = \sum\limits_{n = 0}^\infty \frac{U_a^{(n)}(u, z)}{r^n}\nonumber\\
	&g_{ab} = r^2 \gamma_{ab}(z) + r C_{ab}(u,z)+D_{ab}(u,z)+ \sum\limits_{n = 1}^\infty \frac{g_{ab}^{(n)}(u, z)}{r^n}
\end{align}

Consider the $r$ expansion of the angular part of the metric in $d=6$ as in \eqref{BMS-fall-off-hd}. From the equation of motion it can be shown that, $\partial_{u}C_{ab}(u,z)=0$ and given $\gamma_{ab}(z)$, $C_{ab}(z)$ and $D_{ab}(u,z)$ at $\mathcal{I}^{+}$ the metric can be solved at all order in the bulk. $D_{ab}$ coorsponds to the radiative mode.\footnote{ In \cite{Chowdhury:2022nus}, the authors worked on decompactified sphere, i.e. $\mathbb{S}^{4}\rightarrow \mathbb{R}^{4}$, and so $\gamma_{ab}\rightarrow \delta_{ab}$. However, upon covariantization of the results obtained at the end, one can recover the $\mathbb{S}^{4}$ results.} 
 The above $r$ fall-off \eqref{BMS-fall-off-hd} is preserved by the BMS vector fields, where $\mathfrak{BMS}=\mathfrak{ST}\ltimes\mathrm{Lorentz}$. The components of Supertranslation ($\mathfrak{ST}$) vector fields at all order in r can be written as:
 \begin{align}\label{ST-vector-field-Chandra}
 	&\xi^{u}_{\mathrm{ST}}= f(z)\nonumber\\
 	&\xi^{a}_{\mathrm{ST}}=-\partial_{b}f\int_{\infty}^{r}e^{2\beta}g^{ab}dr^{'}\nonumber\\
 	&\xi^{r}_{\mathrm{ST}}=U^{a}\partial_{a}f-\partial_{a}\xi^{a}.
 \end{align}
Action of supertranslation on $C_{ab}$ and $D_{ab}$ can be written as:
\begin{align}\label{ST-action-Chandra}
	&\delta_{\mathrm{ST}}C_{ab}=-2\bigg[\partial_{a}\partial_{b}f-\frac{1}{4}\delta_{ab}\partial^{2}f\bigg]\nonumber\\
	&\delta_{\mathrm{ST}}D_{ab}=f\partial_{u}D_{ab}+\frac{1}{4}\delta_{ab}\bigg[-\frac{4}{3}\partial_{c}C^{cd}\partial_{d}f-C^{cd}\partial_{c}\partial_{d}f\bigg]+\frac{1}{4}C_{ab}\partial^{2}f-\partial_{c}C_{ab}\partial^{c}f\nonumber\\
	&~~~-\frac{1}{2}\bigg[C_{bc}\partial_{a}\partial^{c}f+C_{ac}\partial_{b}\partial^{c}f\bigg]+\frac{1}{2}\bigg[\partial_{a}C_{bc}\partial^{c}f+\partial_{b}C_{ac}\partial^{c}f\bigg]+\frac{1}{6}\bigg[\partial^{c}C_{bc}\partial^{a}f+\partial^{c}C_{ac}\partial^{b}f\bigg].
\end{align}

  It is important to note that, from the saddle-point analysis and the finiteness of the symplectic structure one expects that the radiative degrees of freedom should scale as $|u|^{-(2+\epsilon)}$  ($\epsilon>0$) at the boundaries of $\mathcal{I}^{+}$. However, as is evident from the \eqref{ST-action-Chandra}, supertranslation action violates this fall-off.  
  
  The news tensor associatated to the radiative degrees of freedom is given by $N_{ab}=\partial_{u}D_{ab}$.  Since, $C_{ab}$ is independent of $u$, redefinition $D_{ab}\rightarrow D_{ab}+ \chi_{ab}$, (where $\chi_{ab}$ is any function constructed from $\gamma_{ab}$ and $C_{ab}$) doesn't change the physical news tensor. 

So, one asks whether there exists a redefinition of the radiative degrees of freedom such that: (1) the redefined field gives same news tensor, (2) $u$ fall-off of this is preserved by supertranslation. It was identified in \cite{Chowdhury:2022nus}, the correct variable for the radiative degrees of freedom in classical theory and hence, correspondingly, the correct graviton mode in the quantized theory  that satisfies the above criteria is not $D_{ab}$, but a non-linear field redefinition given by:
\begin{align}\label{redef-ST}
\tilde{D}^{\mathrm{ST}}_{ab}(u,z)=D_{ab}(u,z)-\frac{1}{4}\delta^{cd}C_{ac}(z)C_{bd}(z)-\frac{1}{16}\delta_{ab}C_{cd}(z)C^{cd}(z)	
\end{align}
Equipped with this redefinition one finds that:
\begin{align}
\delta_{\mathrm{ST}}\tilde{D}^{\mathrm{ST}}_{ab}(u,z)=f(z)\partial_{u}\tilde{D}^{\mathrm{ST}}_{ab}(u,z).	
\end{align}	
Using this redefinition one finds a finite supertranslation charge in $d=6$ for non-linear gravity. The charges can be split into soft and hard part. Note that the soft and hard parts now depend linearly and quadratically on $\tilde{D}_{ab}$ respectively. 

The hard supertranslation charge is given by:
\begin{align}\label{6d-hard}
	\mathcal{Q}^{\mathrm{Hard}}_{\mathrm{ST}}&=\frac{1}{8\pi G_{N}}\int_{\mathcal{I}^{+}}f(z)\mathcal{T}^{(4)}_{uu}(u,z)\nonumber\\&=\frac{1}{8\pi G_{N}}\int_{\mathcal{I}^{+}}f(z)\bigg[\mathcal{T}^{\mathrm{Matter}(4)}_{uu}(u,z)+\frac{1}{4}N^{ab}(u,z)N_{ab}(u,z)\bigg],
\end{align}  
where $N_{ab}=\partial_{u}\tilde{D}^{\mathrm{ST}}_{ab}$ is the News tensor in $d=6$.
Soft Supertranslation Charge is given by:
\begin{align}\label{6d-soft}
	\mathcal{Q}^{\mathrm{Soft}}_{\mathrm{ST}}=\frac{1}{96\pi G_{N}}\int_{\mathcal{I}^{+}}f(z)\partial^{2}\partial^{ab}\tilde{D}^{\mathrm{ST}}_{ab}(u,z)=\frac{1}{96\pi G_{N}}\int_{\mathbb{R}^{4}}f(z)\partial^{2}\partial^{ab}\mathcal{N}^{(0)}_{ab}(z),
\end{align} 
where $\mathcal{N}^{(0)}_{ab}$ is the leading soft mode given by:
\begin{align}
	\mathcal{N}^{(0)}_{ab}(z)=\int_{u}\tilde{D}^{\mathrm{ST}}_{ab}(u,z).
\end{align}
Hence, for the total supertranslation charge we have:
\begin{align}\label{ST-charge-nlin}
	&\mathcal{Q}_{\mathrm{ST}}\nonumber\\
	&=\mathcal{Q}^{\mathrm{Hard}}_{\mathrm{ST}}+\mathcal{Q}^{\mathrm{Soft}}_{\mathrm{ST}}\nonumber\\
							 &=\frac{1}{8\pi G_{N}}\int_{\mathcal{I}^{+}}f(z)\mathcal{T}^{(4)}_{uu}(u,z)+\frac{1}{96\pi G_{N}}\int_{\mathbb{R}^{4}}f(z)\partial^{2}\partial^{ab}\mathcal{N}^{(0)}_{ab}(z)\nonumber\\
							 &=\frac{1}{8\pi G_{N}}\int_{\mathcal{I}^{+}}f(z)\bigg[\mathcal{T}^{\mathrm{Matter}(4)}_{uu}(u,z)+\frac{1}{4}N^{ab}(u,z)N_{ab}(u,z)\bigg]+\frac{1}{96\pi G_{N}}\int_{\mathcal{I}^{+}}f(z)\partial^{2}\partial^{ab}\tilde{D}^{\mathrm{ST}}_{ab}(u,z)
	\end{align}
It is important to note how from this charge \eqref{ST-charge-nlin} one can obtain the linearized gravity charge \eqref{ST-charge-lin} in $d=6$. In the case of linearized gravity the contribution to the energy momentum from the gravitaional news is absent. So replacing   $\tilde{D}^{\mathrm{ST}}_{ab}\rightarrow D_{ab}$ in \eqref{ST-charge-nlin} and decompactifying the $\mathbb{S}^{4}\rightarrow\mathbb{R}^{4}$ in \eqref{ST-charge-lin}, both the charges match.

In \cite{Chowdhury:2022nus}, the authors worked in non-linear gravity and using the charge \eqref{ST-charge-nlin} the connection with the leading single soft graviton theorem was established in the generic $C_{ab}\neq0$ case through a Ward identity of the following form.
\begin{align}
	\bra{\mathrm{out}}[\mathcal{Q}_{\mathrm{ST}},\mathcal{S}]\ket{\mathrm{in}}=0\Leftrightarrow \bra{\mathrm{out}}[\mathcal{Q}^{\mathrm{Soft}}_{\mathrm{ST}},\mathcal{S}]\ket{\mathrm{in}}=-\bra{\mathrm{Out}}[\mathcal{Q}^{\mathrm{Hard}}_{\mathrm{ST}},\mathcal{S}]\ket{\mathrm{in}}
\end{align}
As we discussed previously, the correct graviton mode in this case is not $D_{ab}$ but $\tilde{D}^{\mathrm{ST}}_{ab}$ \cite{Chowdhury:2022nus}. It is important to note that, $C_{ab}$ can be obtained from a scalar potential $\psi$, and supertranslated vacua are labeled by this scalar potential. %Choice of $\tilde{D}^{\mathrm{ST}}_{ab}$ to be the correct variable ensures that there is no preferred Fock vacuum and one can only observe the physical effects of vacuum to vacuum transition.    
%%%%%%%%%%%%%%%%%%%%%%%%%%%%%%%%%%%%%%%%%%%%%%%%%%%%%%
\subsection{Superrotations in Higher Even Dimensions \& Consequences}\label{SR}

So far, we have talked only about supertranslation in higher dimensions. In $d=4$, the $\mathfrak{BMS}$ algebra can be further extended to include superrotations. Also, in $d=4$, the subleading soft graviton theorem follows from the conjectured superrotation symmetry of the quantum gravity $\mathcal{S}$-matrix \cite{Kapec:2014opa,Campiglia:2014yka}. Since the subleading soft theorem holds in any dimension \cite{Sen:2017xjn,Sen:2017nim}, a natural question will be to ask if there is any generalisation of the superrotation symmetry to higher dimensions. As already mentioned, in $d=4$, there are many distinct proposals for superrotations in the sense that all of them are infinite-dimensional extensions of the
Lorentz subalgebra of the $\mathfrak{BMS}~(\mathfrak{ST}\ltimes\mathrm{Lorentz})$ algebra. In $d=4$, Lorentz transformation induces the global conformal transformations on $\mathbb{S}^{2}$. In the Extended BMS ($\mathfrak{EBMS}$) proposal \cite{Barnich:2009se,Barnich:2010eb}, the $\mathrm{Lorentz}$ algebra is extened to include local conformal transformations on $\mathbb{S}^{2}$. In the Generalised BMS ($\mathfrak{GBMS}$) proposal \cite{Campiglia:2015yka}, the $\mathrm{Lorentz}$ algebra is extended to include any area preserving smooth diffeomorphisms of $\mathbb{S}^{2}$. Since, for $d>4$, the corresponding local conformal transformations on $\mathbb{S}^{d-2}$ are finite dimensional, there is no natural generelisation of $\mathfrak{EBMS}$ to higher even dimensions. On the contrary, in higher even dimensions one can attempt to obtain superrotations, and correspondingly a generalisation of $\mathfrak{GBMS}$, from the area-preserving smooth diffeomorphisms on $\mathbb{S}^{d-2}$.  

In the following, we focus mainly on $d=6$, but many of the aspects are expected to have a natural generalisation to any higher even dimensions. We start with the Bondi metric \eqref{bondi-metric}. Inspired by the generalisation of symmetry algebra from $\mathfrak{BMS}$ to $\mathfrak{GBMS}$ in the $d=4$ case, we start by the generalisation of the fall-off conditions chosen for studying the $\mathfrak{BMS}$ algebra. Let us start with the following fall-off conditions \cite{Chowdhury:2022gib}:
\begin{align}\label{GBMS-fall-off-6d}
	&M = \sum\limits_{n = 0}^{\infty} \frac{M^{(n)}(u, z)}{r^n},\quad  
	\beta = \sum\limits_{n = 2}^\infty \frac{\beta^{(n)}(u, z)}{r^n},\quad
	U_a = \sum\limits_{n = 0}^\infty \frac{U_a^{(n)}(u, z)}{r^n}\nonumber\\
	&g_{ab} = r^2 q_{ab}(z) + r C_{ab}(u,z)+D_{ab}(u,z)+\sum\limits_{n = 1}^\infty \frac{g_{ab}^{(n)}(u, z)}{r^n}
\end{align}	
Here, $q_{ab}(z)$ is obtained from any area preserving ($\sqrt{q}=\sqrt{\gamma}$) smooth diffeomorphisms of unit round sphere metric $\gamma_{ab}(z)$. From Einstein's equation  we get $\partial_{u}C_{ab}=-\bar{\mathcal{R}}_{ab}^{TF}$, where $\bar{\mathcal{R}}_{ab}^{TF}$ is the trace-free part of the Ricci tensor corresponding to $q_{ab}$ metric. This implies, $C_{ab}(u,z)=\bar{C}_{ab}(z)+uT_{ab}(z)$, where, $T_{ab}=-\bar{\mathcal{R}}_{ab}^{TF}$. Given the $q_{ab}(z)$, $\bar{C}_{ab}(z)$ and $D_{ab}(u,z)$ metric can be solved at all order in $r$. $D_{ab}(u,z)$ is the dynamical mode.

Connection with the fall-off conditions \eqref{BMS-fall-off-hd} chosen for studying $\mathfrak{BMS}$ algebra must be stressed here. If one restricts to the unit round metric $\gamma_{ab}$ on the $\mathbb{S}^{4}$ i.e. $q_{ab}=\gamma_{ab}$, then $T_{ab}=0$, and $M^{(0)}=-1$, i.e. one essentially recovers \eqref{BMS-fall-off-hd}, and the corresponding symmetry algebra is $\mathfrak{BMS}$. Demanding preservation of the fall-off conditions \eqref{GBMS-fall-off-6d}, one obtains an infinite dimensional extension of the Lorentz subalgebra of original $\mathfrak{BMS}$ algebra, parametrized by any smooth vector field $V^{a}$ on $\mathbb{S}^{4}$. Correspondingly, one gets Generalised-BMS ($\mathfrak{GBMS}$) algebra in $d=6$. Hence, $\mathfrak{GBMS}= \mathfrak{ST}\ltimes \mathrm{Diff}(\mathbb{S}^{4})$. Henceforth, by Superrotation in $d=6$, we shall mean this extension of Lorentz algebra. Superrotation vector fields are given by:
\begin{align}\label{SR-vec-field}
	\xi^u_{\mathrm{SR}} &=  u \alpha(z)\nonumber \\
	\xi^a_{\mathrm{SR}} &= V^a(z) -  u \mathcal{D}_b\alpha(z) \int_r^{\infty}  e^{2\beta(u,r',z)} g^{ab}(u,r',z) dr'\nonumber \\
	\xi^r_{\mathrm{SR}} &= - \frac{r}{4} \big[ \mathcal{D}_a \xi_V^a(u,r,z) - u U^a(u,r,z) \mathcal{D}_a \alpha(z)  \big]~.
\end{align}

Here, $\alpha=\frac{1}{4}\mathcal{D}_{a}V^{a}$. Action of the superrotation on $\bar{C}_{ab}$, $T_{ab}$ and $D_{ab}$ can be written as:
\begin{align}\label{SR-action-hd}
	&\delta_{\mathrm{SR}} \bar C_{ab} = \mathcal{L}_V \bar C_{ab} - \alpha \bar C_{ab}\nonumber \\
	&\delta_{\mathrm{SR}} T_{ab} = \mathcal{L}_V T_{ab}  - 2  \big( \mathcal{D}_a \mathcal{D}_b \alpha \big)^{TF}\nonumber \\
	&\delta_{\mathrm{SR}} D_{ab} = u \alpha \partial_u D_{ab} + \mathcal{L}_V  D_{ab}\nonumber \\
	&\quad\quad\quad\quad+ u \Bigg\{ \frac{1}{4} \mathcal{D}^2 \alpha  C_{ab} -  U_{(a}^{(0)} \mathcal{D}_{b)} \alpha + \frac{1}{2} q_{c(a} \mathcal{D}_{b)} \big( C^{cd} \mathcal{D}_d \alpha \big) - C_{c(a} \mathcal{D}_{b)} \mathcal{D}^c \alpha \nonumber\\
	&\qquad\quad\quad - \mathcal{D}^c \alpha \mathcal{D}_c C_{ab}  + \frac{1}{2}q_{ab} U^{(0)c} \mathcal{D}_c \alpha - \frac{1}{4}q_{ab} \mathcal{D}_c (C^{cd} \mathcal{D}_d \alpha)  \Bigg\} 
\end{align}

 Similar to \cite{Chowdhury:2022nus}, we shall work on the decompactified sphere ($\mathbb{R}^{4}$). Borrowing from the terminology used in $d=4$, we call the case of $q_{ab}=\gamma_{ab}$ metric on  $\mathbb{S}^{4}$ (or $\delta_{ab}$ metric on $\mathbb{R}^{4}$) as Bondi frame. In the Bondi frame, $T_{ab}=0$ and $C_{ab}=\bar{C}_{ab}$ and hence, the superrotation action \eqref{SR-action-hd} takes a simpler form. Now, superrotation action takes away from the Bondi frame i.e. $\delta_{\mathrm{SR}}T_{ab}\neq0$, even starting from Bondi frame where $T_{ab}=0$.

 Due to the generalisation $r$ fall-off condition from \eqref{BMS-fall-off-hd} to \eqref{GBMS-fall-off-6d}, there arises a need for further field redefinition of radiative degrees of freedom, such that the $u$ fall-off at the boundaries of the $\mathcal{I}^{+}$ is maintained. This generalisation should capture the information of non-zero $T_{ab}$, but should smoothly reproduce the redefinition \eqref{redef-ST} in the Bondi case ($T_{ab}=0$, $C_{ab}=\bar{C}_{ab}$). We shall look at the effect of going linearly away from the Bondi frame. In this case, a natural generalisation of field redefinition becomes:
\begin{align}\label{redef-SR}
	\tilde D_{ab} &= D_{ab} - \frac{1}{4} q^{mn} \bar C_{am} \bar C_{bn} - \frac{1}{16} q_{ab}\bar C_{mn} \bar C^{mn}\nonumber \\
	&\quad - u \Big[ \frac{1}{4} q^{mn} (\bar C_{am} T_{bn} + T_{am} \bar C_{bn}) + \frac{1}{8} q_{ab} T_{mn} \bar C^{mn} \Big] + O(T^2).
\end{align}
Note that supertranslation and superrotation action on this redefined radiative field can be written as:
\begin{align}\label{Dtilvariations}
	\delta_{\mathrm{ST}} \tilde D_{ab}  &= f \partial_u \tilde D_{ab}~ \\
	\delta_{\mathrm{SR}} \tilde D_{ab} &= \mathcal{L}_V \tilde D_{ab} + u \alpha \partial_u \tilde D_{ab}~
\end{align}
Thus, the $u$ fall-offs are not violated by supertranslation or superotation action starting from a Bondi frame.

In \cite{Chowdhury:2022gib}, conserved charge corresponding to superrotation symmetry in the Bondi frame was obtained. Superrotation hard charge was derived from the energy momentum tensor as follows:
\begin{align}\label{SR-Hard-Stress}
	\mathcal{Q}^{\mathrm{Hard}}_{\mathrm{SR}}=\frac{1}{8\pi G_{N}}\int_{\mathcal{I}^{+}}\bigg[u\alpha(z)\mathcal{T}^{(4)}_{uu}(u,z) + V^a(z) \mathcal{T}^{(4)}_{ua}(u,z)\bigg].
\end{align}
Hence, for pure gravity superrotation hard charge was obtained to be:
\begin{align}\label{SR-Hard}
	\mathcal{Q}^{\mathrm{Hard}}_{\mathrm{SR}} = \frac{1}{32\pi G_{N}}  \int_{\mathcal{I}^{+}}\ N^{ab} \Big( \mathcal{L}_V \tilde D_{ab} + u \alpha N_{ab} \Big).
\end{align} 
where, $N_{ab}=\partial_{u}\tilde{D}_{ab}$ is the news tensor.
In \cite{Chowdhury:2022nus}, the following superrotation soft charge was proposed for any generic Bondi frame ($\bar{C}_{ab}\neq 0$):
\begin{align}\label{SR-soft}
	\mathcal{Q}^{\mathrm{Soft}}_{\mathrm{SR}}=  \frac{1}{128\pi G_{N}}\int_{\mathcal{I}^{+}} u V^b(x) \Big[ \partial^4 \partial^a \tilde D_{ab} - \frac{4}{3} \partial_b \partial^2 \partial^{ef} \tilde D_{ef}   \Big]  \nonumber\\
	+ \frac{1}{96\pi G_{N}} \int_{\mathcal{I}^{+}} (\mathcal{L}_V \bar C_{ab} - \alpha \bar C_{ab}) \partial^a \partial^m \tilde D^b_m
\end{align}
The correctness of the soft charge is tested by the fact that they produce correct action on the Kinametic data ($\bar{C}_{ab}, T_{ab}$) in Bondi frame. This soft charge is further justified by the fact that they reproduce the correct subleading soft graviton theorem in the quantum theory.

Subleading Single Soft Graviton Theorem \cite{Cachazo:2014fwa} in any dimension including $d=6$ is given by \eqref{subleading-ST}. Now, $\mathfrak{GBMS}$ compatible fall-off \eqref{GBMS-fall-off-6d} implies that the correct graviton mode to quantize is $\tilde{D}_{ab}$. We choose the vacua to be labelled by the simultaneous eigenstates of $\bar{C}_{ab}$ and $T_{ab}$. Ordinary Fock vacuum is identified as $\ket{0}=\ket{\bar{C}_{ab}=0, T_{ab}=0}$.

Next, we consider a scenario of a massless scalar field coupled with gravity and consider the implication of superrotation symmetry to the $\mathcal{S}$-matrix of this theory. In this theory, soft charge is given by \eqref{SR-soft} and the hard charge is obtained from the corresponding stress energy tensor of scalar using \eqref{SR-Hard-Stress}. 

Next one looks at the Ward identity:
\begin{align}
	\bra{\mathrm{out}}[\mathcal{Q}_{\mathrm{SR}},\mathcal{S}]\ket{\mathrm{in}}=0\Leftrightarrow \bra{\mathrm{out}}[\mathcal{Q}^{\mathrm{Soft}}_{\mathrm{SR}},\mathcal{S}]\ket{\mathrm{in}}=-\bra{\mathrm{Out}}[\mathcal{Q}^{\mathrm{Hard}}_{\mathrm{SR}},\mathcal{S}]\ket{\mathrm{in}}
\end{align}
In \cite{Chowdhury:2022gib}, it was found that this identity can be obtained as a consequence of the subleading soft graviton theorem \eqref{subleading-ST} in $d=6$. It is important to note that, the action of the second term in \eqref{SR-soft} on the ordinary Fock vacuum is zero due to the normal ordering of the operators chosen in \cite{Chowdhury:2022gib}. Reproduction of the subleading single soft graviton theorem justifies the correctness of the proposed soft charge in \eqref{SR-soft}.

For a very recent and rigorous study on the phase space of gravity in six-dimensional asymptotically flat spacetime, we refer the reader to \cite{Capone:2023roc}. For a study of general superrotation compatible kinematic space of gravity in generic higher even dimensions, we refer the reader to \cite{Capone:2021ouo}. 

 % % % % % % % % % % % % % % % % % % % % % % % % % % % % % % % % % % % % % % % % % % % % % % % % % % % % % % % % % % % %
\section{Summary and Open Issues}\label{summary}
Let us summarize what we have discussed so far. We started with a discussion of early results on asymptotic symmetries in $d=4$ and higher. In particular, we stated that how ealy results set no-go conditions on the existance of non-trivial asymptotic symmetries in higher even dimensions. Then we discussed new insights gained from certain results regarding quantum gravity $\mathcal{S}$-matrix and the need for the existence of supertranslation in the higher even dimensions.

 We first discussed the consequences of Supertranslation in linearized gravity at the classical level in $d=2m+2$ dimensions and discussed the conserved asymptotic charge. We stressed how earlier no-go conditions could be bypassed by imposing certain late and early time behaviour on the metric components. Then we specialised to $d=6$ and discussed the consequences of supertranslation symmetry both at the level of linear as well as non-linear theory. An important lesson from the non-linear theory is that, to make the physically necessary fall-off of the radiative degrees of freedom at the boundaries of null infinity supertranslation compatible, a non-linear field redefinition of the radiative degrees of freedom is needed. 

Next, we discussed the consequences of extending the symmetry to include superrotation. The lesson that we get is: to have the physically necessary fall-off of the radiative degrees of freedom both Supertranslation and superrotation compatible it becomes necessary to do a further non-linear redefinition of radiative degrees of freedom. We discussed the conserved asymptotic charge that one gets from the Superrotation. Next, we briefly discussed how by elevation of this symmetry to the symmetry of quantum gravity $\mathcal{S}$-matrix a connction with the subleading soft graviton theorem can be made.

Many aspects remain open-ended. In $d=4$, we now understand how the double soft graviton factorisations are connected to asymptotic symmetries \cite{Distler:2018rwu,Campiglia:2021bap}. A similar derivation in higher dimensions is not yet known. Also, in $d=4$ for the case, when there are massive particles in the external states, one knows how to build the connection between the single soft graviton theorems and asymptotic symmetries \cite{Campiglia:2015kxa}. Similar derivation in higher even dimensions is yet to be done.

Soft graviton theorems hold in all dimensions. However, the study of asymptotic symmetries in odd dimensions and their possible connection to soft theorems still remain a largely open issue. However, important progress has been made in recent works \cite{Fuentealba:2022yqt}.

In this article,  we have focussed on spacetimes with higher non-compact dimensions. For study of higher dimensional spacetimes with compact extra dimensions we refer the readers to \cite{Ferko:2021bym}.

\section*{Acknowledgement}
I thank Alok Laddha for his constant encouragement for writing this review article and his crucial comments on the first draft. I learned many aspects of the topic of asymptotic symmetries through the discussions and collaborations with Alok Laddha, Anupam AH, Chandramouli Chowdhury, Ankit Aggarwal, Aniket Khairnar, Krishnendu Roy, Miguel Campiglia, Amitabh Virmani, Arnab Priya Saha, Nishant Agarwal, Amit Suthar, Shamim Akhtar at different stages of the past few years. I would like to thank Shrihari Goplakrishnan, V. Ravindran, Sujay Ashok, and various other members of the IMSc high energy physics group for their various conceptual questions, which encouraged me to investigate various minute aspects of the subject.

\newpage	
	\bibliography{review} 

\providecommand{\href}[2]{#2}\begingroup\raggedright\begin{thebibliography}{10}

\bibitem{He:2014laa}
T.~He, V.~Lysov, P.~Mitra and A.~Strominger, \emph{{BMS supertranslations and
  Weinberg\textquoteright{}s soft graviton theorem}},
  \href{https://doi.org/10.1007/JHEP05(2015)151}{\emph{JHEP} {\bfseries 05}
  (2015) 151}, [\href{https://arxiv.org/abs/1401.7026}{{\ttfamily 1401.7026}}].

\bibitem{Strominger:2017zoo}
A.~Strominger, \emph{{Lectures on the Infrared Structure of Gravity and Gauge
  Theory}},  \href{https://arxiv.org/abs/1703.05448}{{\ttfamily 1703.05448}}.

\bibitem{Penedones:2016voo}
J.~Penedones, \emph{{TASI lectures on AdS/CFT}},  in \emph{{Theoretical
  Advanced Study Institute in Elementary Particle Physics}: {New Frontiers in
  Fields and Strings}}, pp.~75--136, 2017,
  \href{https://arxiv.org/abs/1608.04948}{{\ttfamily 1608.04948}},
  \href{https://doi.org/10.1142/9789813149441_0002}{DOI}.

\bibitem{Hollands:2004ac}
S.~Hollands and R.~M. Wald, \emph{{Conformal null infinity does not exist for
  radiating solutions in odd spacetime dimensions}},
  \href{https://doi.org/10.1088/0264-9381/21/22/008}{\emph{Class. Quant. Grav.}
  {\bfseries 21} (2004) 5139--5146},
  [\href{https://arxiv.org/abs/gr-qc/0407014}{{\ttfamily gr-qc/0407014}}].

\bibitem{Bondi:1962px}
H.~Bondi, M.~G.~J. van~der Burg and A.~W.~K. Metzner, \emph{{Gravitational
  waves in general relativity. 7. Waves from axisymmetric isolated systems}},
  \href{https://doi.org/10.1098/rspa.1962.0161}{\emph{Proc. Roy. Soc. Lond. A}
  {\bfseries 269} (1962) 21--52}.

\bibitem{Sachs:1962wk}
R.~K. Sachs, \emph{{Gravitational waves in general relativity. 8. Waves in
  asymptotically flat space-times}},
  \href{https://doi.org/10.1098/rspa.1962.0206}{\emph{Proc. Roy. Soc. Lond. A}
  {\bfseries 270} (1962) 103--126}.

\bibitem{Barnich:2009se}
G.~Barnich and C.~Troessaert, \emph{{Symmetries of asymptotically flat 4
  dimensional spacetimes at null infinity revisited}},
  \href{https://doi.org/10.1103/PhysRevLett.105.111103}{\emph{Phys. Rev. Lett.}
  {\bfseries 105} (2010) 111103},
  [\href{https://arxiv.org/abs/0909.2617}{{\ttfamily 0909.2617}}].

\bibitem{Barnich:2010eb}
G.~Barnich and C.~Troessaert, \emph{{Aspects of the BMS/CFT correspondence}},
  \href{https://doi.org/10.1007/JHEP05(2010)062}{\emph{JHEP} {\bfseries 05}
  (2010) 062}, [\href{https://arxiv.org/abs/1001.1541}{{\ttfamily 1001.1541}}].

\bibitem{Campiglia:2015yka}
M.~Campiglia and A.~Laddha, \emph{{New symmetries for the Gravitational
  S-matrix}}, \href{https://doi.org/10.1007/JHEP04(2015)076}{\emph{JHEP}
  {\bfseries 04} (2015) 076},
  [\href{https://arxiv.org/abs/1502.02318}{{\ttfamily 1502.02318}}].

\bibitem{Hollands:2016oma}
S.~Hollands, A.~Ishibashi and R.~M. Wald, \emph{{BMS Supertranslations and
  Memory in Four and Higher Dimensions}},
  \href{https://doi.org/10.1088/1361-6382/aa777a}{\emph{Class. Quant. Grav.}
  {\bfseries 34} (2017) 155005},
  [\href{https://arxiv.org/abs/1612.03290}{{\ttfamily 1612.03290}}].

\bibitem{Hollands:2003ie}
S.~Hollands and A.~Ishibashi, \emph{{Asymptotic flatness and Bondi energy in
  higher dimensional gravity}},
  \href{https://doi.org/10.1063/1.1829152}{\emph{J. Math. Phys.} {\bfseries 46}
  (2005) 022503}, [\href{https://arxiv.org/abs/gr-qc/0304054}{{\ttfamily
  gr-qc/0304054}}].

\bibitem{Tanabe:2011es}
K.~Tanabe, S.~Kinoshita and T.~Shiromizu, \emph{{Asymptotic flatness at null
  infinity in arbitrary dimensions}},
  \href{https://doi.org/10.1103/PhysRevD.84.044055}{\emph{Phys. Rev. D}
  {\bfseries 84} (2011) 044055},
  [\href{https://arxiv.org/abs/1104.0303}{{\ttfamily 1104.0303}}].

\bibitem{Weinberg:1965nx}
S.~Weinberg, \emph{{Infrared photons and gravitons}},
  \href{https://doi.org/10.1103/PhysRev.140.B516}{\emph{Phys. Rev.} {\bfseries
  140} (1965) B516--B524}.

\bibitem{Cachazo:2014fwa}
F.~Cachazo and A.~Strominger, \emph{{Evidence for a New Soft Graviton
  Theorem}},  \href{https://arxiv.org/abs/1404.4091}{{\ttfamily 1404.4091}}.

\bibitem{Sen:2017xjn}
A.~Sen, \emph{{Soft Theorems in Superstring Theory}},
  \href{https://doi.org/10.1007/JHEP06(2017)113}{\emph{JHEP} {\bfseries 06}
  (2017) 113}, [\href{https://arxiv.org/abs/1702.03934}{{\ttfamily
  1702.03934}}].

\bibitem{Sen:2017nim}
A.~Sen, \emph{{Subleading Soft Graviton Theorem for Loop Amplitudes}},
  \href{https://doi.org/10.1007/JHEP11(2017)123}{\emph{JHEP} {\bfseries 11}
  (2017) 123}, [\href{https://arxiv.org/abs/1703.00024}{{\ttfamily
  1703.00024}}].

\bibitem{Laddha:2017ygw}
A.~Laddha and A.~Sen, \emph{{Sub-subleading Soft Graviton Theorem in Generic
  Theories of Quantum Gravity}},
  \href{https://doi.org/10.1007/JHEP10(2017)065}{\emph{JHEP} {\bfseries 10}
  (2017) 065}, [\href{https://arxiv.org/abs/1706.00759}{{\ttfamily
  1706.00759}}].

\bibitem{Chakrabarti:2017ltl}
S.~Chakrabarti, S.~P. Kashyap, B.~Sahoo, A.~Sen and M.~Verma, \emph{{Subleading
  Soft Theorem for Multiple Soft Gravitons}},
  \href{https://doi.org/10.1007/JHEP12(2017)150}{\emph{JHEP} {\bfseries 12}
  (2017) 150}, [\href{https://arxiv.org/abs/1707.06803}{{\ttfamily
  1707.06803}}].

\bibitem{Sahoo:2018lxl}
B.~Sahoo and A.~Sen, \emph{{Classical and Quantum Results on Logarithmic Terms
  in the Soft Theorem in Four Dimensions}},
  \href{https://doi.org/10.1007/JHEP02(2019)086}{\emph{JHEP} {\bfseries 02}
  (2019) 086}, [\href{https://arxiv.org/abs/1808.03288}{{\ttfamily
  1808.03288}}].

\bibitem{Strominger:2013jfa}
A.~Strominger, \emph{{On BMS Invariance of Gravitational Scattering}},
  \href{https://doi.org/10.1007/JHEP07(2014)152}{\emph{JHEP} {\bfseries 07}
  (2014) 152}, [\href{https://arxiv.org/abs/1312.2229}{{\ttfamily 1312.2229}}].

\bibitem{Strominger:2014pwa}
A.~Strominger and A.~Zhiboedov, \emph{{Gravitational Memory, BMS
  Supertranslations and Soft Theorems}},
  \href{https://doi.org/10.1007/JHEP01(2016)086}{\emph{JHEP} {\bfseries 01}
  (2016) 086}, [\href{https://arxiv.org/abs/1411.5745}{{\ttfamily 1411.5745}}].

\bibitem{Kapec:2014opa}
D.~Kapec, V.~Lysov, S.~Pasterski and A.~Strominger, \emph{{Semiclassical
  Virasoro symmetry of the quantum gravity $ \mathcal{S}$-matrix}},
  \href{https://doi.org/10.1007/JHEP08(2014)058}{\emph{JHEP} {\bfseries 08}
  (2014) 058}, [\href{https://arxiv.org/abs/1406.3312}{{\ttfamily 1406.3312}}].

\bibitem{Campiglia:2014yka}
M.~Campiglia and A.~Laddha, \emph{{Asymptotic symmetries and subleading soft
  graviton theorem}},
  \href{https://doi.org/10.1103/PhysRevD.90.124028}{\emph{Phys. Rev. D}
  {\bfseries 90} (2014) 124028},
  [\href{https://arxiv.org/abs/1408.2228}{{\ttfamily 1408.2228}}].

\bibitem{Pasterski:2015tva}
S.~Pasterski, A.~Strominger and A.~Zhiboedov, \emph{{New Gravitational
  Memories}}, \href{https://doi.org/10.1007/JHEP12(2016)053}{\emph{JHEP}
  {\bfseries 12} (2016) 053},
  [\href{https://arxiv.org/abs/1502.06120}{{\ttfamily 1502.06120}}].

\bibitem{Kapec:2015vwa}
D.~Kapec, V.~Lysov, S.~Pasterski and A.~Strominger, \emph{{Higher-dimensional
  supertranslations and Weinberg\textquoteright{}s soft graviton theorem}},
  \href{https://doi.org/10.4310/AMSA.2017.v2.n1.a2}{\emph{Ann. Math. Sci.
  Appl.} {\bfseries 02} (2017) 69--94},
  [\href{https://arxiv.org/abs/1502.07644}{{\ttfamily 1502.07644}}].

\bibitem{Colferai:2020rte}
D.~Colferai and S.~Lionetti, \emph{{Asymptotic symmetries and the subleading
  soft graviton theorem in higher dimensions}},
  \href{https://doi.org/10.1103/PhysRevD.104.064010}{\emph{Phys. Rev. D}
  {\bfseries 104} (2021) 064010},
  [\href{https://arxiv.org/abs/2005.03439}{{\ttfamily 2005.03439}}].

\bibitem{Chowdhury:2022gib}
C.~Chowdhury, A.~A. H. and A.~Kundu, \emph{{Generalized BMS algebra in higher
  even dimensions}},
  \href{https://doi.org/10.1103/PhysRevD.106.126025}{\emph{Phys. Rev. D}
  {\bfseries 106} (2022) 126025},
  [\href{https://arxiv.org/abs/2209.06839}{{\ttfamily 2209.06839}}].

\bibitem{Capone:2023roc}
F.~Capone, P.~Mitra, A.~Poole and B.~Tomova, \emph{{Phase Space Renormalization
  and Finite BMS Charges in Six Dimensions}},
  \href{https://arxiv.org/abs/2304.09330}{{\ttfamily 2304.09330}}.

\bibitem{Aggarwal:2018ilg}
A.~Aggarwal, \emph{{Supertranslations in Higher Dimensions Revisited}},
  \href{https://doi.org/10.1103/PhysRevD.99.026015}{\emph{Phys. Rev. D}
  {\bfseries 99} (2019) 026015},
  [\href{https://arxiv.org/abs/1811.00093}{{\ttfamily 1811.00093}}].

\bibitem{Compere:2018aar}
G.~Comp\`ere and A.~Fiorucci, \emph{{Advanced Lectures on General Relativity}},
   \href{https://arxiv.org/abs/1801.07064}{{\ttfamily 1801.07064}}.

\bibitem{Chowdhury:2022nus}
C.~Chowdhury, R.~Mishra and S.~G. Prabhu, \emph{{The Asymptotic Structure of
  Gravity in Higher Even Dimensions}},
  \href{https://arxiv.org/abs/2201.07813}{{\ttfamily 2201.07813}}.

\bibitem{Christodoulou:1993uv}
D.~Christodoulou and S.~Klainerman, \emph{{The Global nonlinear stability of
  the Minkowski space}}, .

\bibitem{Capone:2021ouo}
F.~Capone, \emph{{General null asymptotics and superrotation-compatible
  configuration spaces in $d\ge4$}},
  \href{https://doi.org/10.1007/JHEP10(2021)158}{\emph{JHEP} {\bfseries 10}
  (2021) 158}, [\href{https://arxiv.org/abs/2108.01203}{{\ttfamily
  2108.01203}}].

\bibitem{Distler:2018rwu}
J.~Distler, R.~Flauger and B.~Horn, \emph{{Double-soft graviton amplitudes and
  the extended BMS charge algebra}},
  \href{https://doi.org/10.1007/JHEP08(2019)021}{\emph{JHEP} {\bfseries 08}
  (2019) 021}, [\href{https://arxiv.org/abs/1808.09965}{{\ttfamily
  1808.09965}}].

\bibitem{Campiglia:2021bap}
M.~Campiglia and A.~Laddha, \emph{{BMS Algebra, Double Soft Theorems, and All
  That}},  \href{https://arxiv.org/abs/2106.14717}{{\ttfamily 2106.14717}}.

\bibitem{Campiglia:2015kxa}
M.~Campiglia and A.~Laddha, \emph{{Asymptotic symmetries of gravity and soft
  theorems for massive particles}},
  \href{https://doi.org/10.1007/JHEP12(2015)094}{\emph{JHEP} {\bfseries 12}
  (2015) 094}, [\href{https://arxiv.org/abs/1509.01406}{{\ttfamily
  1509.01406}}].

\bibitem{Fuentealba:2022yqt}
O.~Fuentealba, M.~Henneaux, J.~Matulich and C.~Troessaert, \emph{{Asymptotic
  structure of the gravitational field in five spacetime dimensions:
  Hamiltonian analysis}},
  \href{https://doi.org/10.1007/JHEP07(2022)149}{\emph{JHEP} {\bfseries 07}
  (2022) 149}, [\href{https://arxiv.org/abs/2206.04972}{{\ttfamily
  2206.04972}}].

\bibitem{Ferko:2021bym}
C.~Ferko, G.~Satishchandran and S.~Sethi, \emph{{Gravitational memory and
  compact extra dimensions}},
  \href{https://doi.org/10.1103/PhysRevD.105.024072}{\emph{Phys. Rev. D}
  {\bfseries 105} (2022) 024072},
  [\href{https://arxiv.org/abs/2109.11599}{{\ttfamily 2109.11599}}].

\end{thebibliography}\endgroup
	\bibliographystyle{JHEP}
	
\end{document}